\documentclass[conference]{IEEEtran}
\IEEEoverridecommandlockouts
\usepackage{cite}
\usepackage{amsmath,amssymb,amsfonts}
\usepackage{algorithmic}
\usepackage{graphicx}
\usepackage{textcomp}
\usepackage{xcolor}
\usepackage{multirow}
\usepackage{booktabs}
\usepackage{multirow}
\usepackage{hyperref}

\def\BibTeX{{\rm B\kern-.05em{\sc i\kern-.025em b}\kern-.08em
    T\kern-.1667em\lower.7ex\hbox{E}\kern-.125emX}}
\begin{document}

\title{Generation of realistic cardiac ultrasound sequences with ground truth motion and speckle decorrelation
\thanks{This work was supported by the NSERC Canada Graduate Scholarships-Doctoral programs, the FRQNT Doctoral Training Scholarships, the French National Research Agency (LABEX PRIMES [ANR-11-LABX-0063], and ORCHID [ANR-22-CE45-0029-01] project).}
}


\DeclareRobustCommand{\IEEEauthorrefmark}[1]{\smash{\textsuperscript{\footnotesize #1}}}

\author{\IEEEauthorblockN{Thierry Judge\IEEEauthorrefmark{1,2},
Nicolas Duchateau\IEEEauthorrefmark{1,3},  Khuram Faraz\IEEEauthorrefmark{1}, Pierre-Marc Jodoin\IEEEauthorrefmark{2} and
Olivier Bernard\IEEEauthorrefmark{1}}
\IEEEauthorblockA{
\IEEEauthorrefmark{1} INSA, Universite Claude Bernard Lyon 1, CNRS UMR 5220, Inserm U1206, CREATIS, Villeurbanne, France. \\
\IEEEauthorrefmark{2} Department of Computer Science, University of Sherbrooke, Sherbrooke, QC, Canada \\
\IEEEauthorrefmark{3}Institut Universitaire de France (IUF) \\
thierry.judge@usherbrooke.ca}}
\maketitle

\begin{abstract}
Simulated ultrasound image sequences are key for training and validating machine learning algorithms for left ventricular strain estimation. Several simulation pipelines have been proposed to generate sequences with corresponding ground truth motion, but they suffer from limited realism as they do not consider speckle decorrelation. In this work, we address this limitation by proposing an improved simulation framework that explicitly accounts for speckle decorrelation. Our method builds on an existing ultrasound simulation pipeline by incorporating a dynamic model of speckle variation. Starting from real ultrasound sequences and myocardial segmentations, we generate meshes that guide image formation. Instead of applying a fixed ratio of myocardial and background scatterers, we introduce a coherence map that adapts locally over time. This map is derived from correlation values measured directly from the real ultrasound data, ensuring that simulated sequences capture the characteristic temporal changes observed in practice. We evaluated the realism of our approach using ultrasound data from 98 patients in the CAMUS database. Performance was assessed by comparing correlation curves from real and simulated images. The proposed method achieved lower mean absolute error compared to the baseline pipeline, indicating that it more faithfully reproduces the decorrelation behavior seen in clinical data.
\end{abstract}


\section{Introduction}

Ultrasound imaging is a key modality in cardiac diagnosis due to its real-time capabilities, portability, and low cost. However, analyzing ultrasound data remains time-consuming and often requires expert interpretation. In recent years, deep learning has shown strong performance in segmentation tasks for ultrasound imaging, enabling automated analysis of anatomical structures. Despite this progress, deep learning has seen limited adoption in motion tracking, where speckle tracking echocardiography (STE) stands as the clinical standard.

STE relies on the assumption that speckle patterns remain stable over time, allowing them to be used as natural markers for motion tracking. However, this assumption often breaks down in practice due to speckle decorrelation, a well-known phenomenon in ultrasound imaging. As a result, STE algorithms require extensive vendor-specific tuning and regularization to maintain performance.

Recently, deep learning-based methods for motion estimation have advanced significantly, particularly in the fields of dense optical flow and point tracking~\cite{Teed_raft,karaev_cotracker}. Several adaptations of these methods for ultrasound imaging have been proposed \cite{evain_camusyn, azad_echotracker}. These approaches are typically trained using either reference data from STE tools or ultrasound simulations. As STE-derived references are difficult to obtain at scale and may carry inherent biases, many works have proposed 
to generate realistic ultrasound simulation videos with reference motion. Evain \textit{et al.} proposed generating videos with motion derived from 2D+t segmentation \cite{evain_camusyn}. Burman \textit{et al.} proposed using a heart model to generate realistic motion from cardiovascular mechanics\cite{burman_dataset}. A common limitation in previous studies is the lack of texture-varying realism due to the absence of speckle-decorrelation modeling. In real acquisitions, speckle patterns are not constant in time and space. Ignoring this variability can impair the performance and generalization of deep learning methods, particularly when applied to lower-quality images from routine clinical practice.

To address this gap, we introduce a simulation pipeline that explicitly models spatiotemporal speckle decorrelation. Building on \cite{evain_camusyn}, we adjust the decorrelation level to match that observed in clinical acquisitions, generating sequences whose speckle dynamics closely resemble real data and are therefore better suited for training and evaluating motion-estimation networks.

\section{Method}

\begin{figure}
    \centering
    \includegraphics[width=1\linewidth]{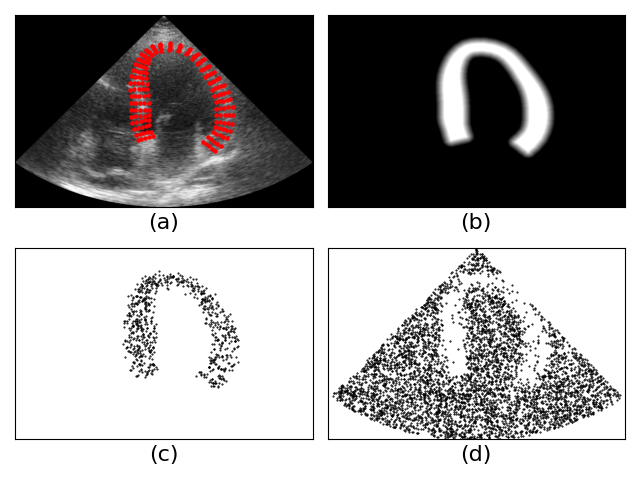}
    \caption{Illustration of the different key elements used in strategy 1: (a) Mesh with $l=36$ and $r=5$, (b) Static coherence map for one frame, (c) and (d) Myocardial and background scatterers for one frame.}
    \label{fig:s1}
\end{figure}

\begin{figure*}
    \centering
    \includegraphics[width=1\linewidth]{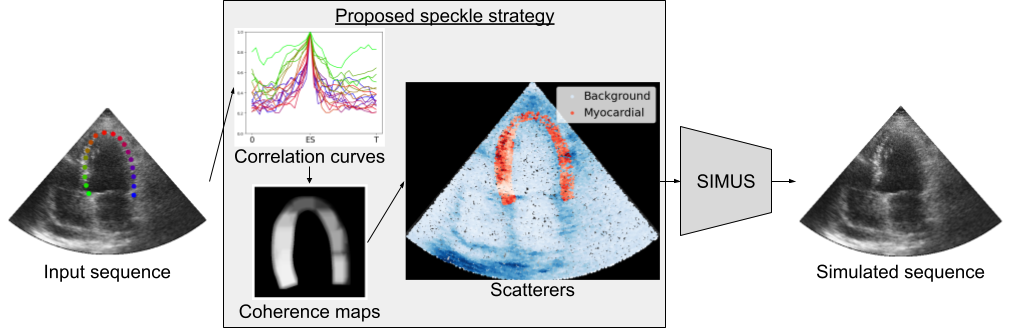}
    \caption{Illustration of the proposed simulation pipeline. The correlation is first measured in the first image before being used to define the dynamics of the myocardial and background scatterers, creating a more realistic simulation. For clarity, only a subset of points in the mesh are illustrated in the input sequence and correlation curves.}
    \label{fig:pipeline}
\end{figure*}

The goal of the simulation pipeline is to create simulated videos with reference motion. These simulations are based on real videos and approximations of the true motion. The output sequence looks similar in appearance to the input one, but its speckle patterns now move according to the ground truth motion introduced by the pipeline. We start by introducing the strategy used by Evain \textit{et al.} \cite{evain_camusyn}. 

\subsection{Strategy 1: Base simulation pipeline}

The input of the pipeline is a video, $\mathbf{V} \in [0,1]^{T \times H \times W}$,  with $T$ frames of height $H$ and width $W$ and a mesh $M \in \mathbb{R}^{T \times l \times r \times 2}$ representing the decomposition of the myocardial wall segmentation and roughly corresponding to the approximate global motion of the myocardium in the video, where $l$ and $r$ are the number of longitudinal points and radial points in the mesh respectively (Fig. \ref{fig:s1} (a)). From the mesh, a mask can be obtained with the bounding polygon of the mesh. 

The pipeline revolves around the SIMUS simulation~\cite{garcia_simus}. For each frame, the input to the simulator is a set of scatterers characterized by spatial coordinates and backscatter coefficients (BSC). Each scatterer $i$ is defined by a lateral and depth position at time $t$, $x_t^i,z_t^i$, and a backscatter reflection coefficient $\text{BSC}_t^i$. To obtain the BSC at time $t$, the value is extracted from the real video with: 
\begin{equation}
    \text{BSC}_t^i = (\mathbf{V}_{t, x_t^i,z_t^i})^\gamma \cdot \epsilon, \quad \epsilon \sim \mathcal{N}(0,1),
    \label{eq:bsc}
\end{equation}
where $\gamma$ is the gamma compression constant and $\epsilon$ is a random variable iid of standard normal distribution.

In \cite{evain_camusyn}, two types of scatterers are defined: myocardial scatterers that follow the movement of the myocardium and background scatterers that complete the image. To define the two types of scatterers, a static coherence map, $\mathbf{C}^s \in [0,1]^{T \times H \times W}$, is obtained by multiplying the myocardial mask by a probability value $p \in [0, 1]$ (Fig. \ref{fig:s1} (b)). To ensure a smooth transition between the myocardium and background, the value decreases linearly from the mask edge to 0 in the background. 

Using the end-systolic (ES) frame as a reference, as most of the myocardium is usually in the sector, scatterers are distributed uniformly across the ultrasound sector with a density of 5 per square wavelength. Myocardial scatterers are selected randomly using the probability given by the coherence map at the position of each scatterer. The BSC of myocardial scatterers is fixed, and their position is updated at each time frame using their relative position in the mesh. The background scatterers have a fixed position at each time step. Their BSC value is updated for each frame with Eq.\ref{eq:bsc}. Finally, a certain number of background scatterers, defined by the coherence map, are randomly selected in the myocardium over the cardiac cycle to create a more realistic speckle pattern. An illustration of myocardial and background scatterers is shown in Figs.~\ref{fig:s1} (c) and (d) with $20\%$ of the scatterers visible for clarity.

Given the combined list of myocardial and background scatterers at each frame, as well as probe parameters, the SIMUS simulator~\cite{garcia_simus} is used to obtain RF signals which are then beamformed to obtain B-mode images.

\subsection{Strategy 2: Integrating speckle correlation}

The previously described simulation pipeline has a constant ratio of coherent and incoherent scatterers. In practice, this leads to coherent speckle patterns across the simulated video. To fix this issue, we propose the method illustrated in Fig.~\ref{fig:pipeline} that dynamically adjusts the ratio of myocardial and background scatterers according to the real video. In order to quantify the level of speckle correlation with respect to the original video and to adjust the simulation, it is necessary to measure the speckle correlation in the real input video. 

To do so, for all the points in the mesh, we compute the correlation between the end-systole instant and all other instants of the input video. Concretely, we define a $25 \times 25$ square window centered at every point in the ES frame. This window is the reference for all other time frames. Then, for every point at every instant $t$ in the video, we compute the normalized 2D correlation between a window centered at the new point's position and the reference window. The maximum value of the resulting 2D matrix is taken as the correlation value for the point at time $t$. Taking the maximum value of the window accounts for possible errors in the mesh, as the real position of the point might be different.  Once the correlation is computed for all points in the mesh at every time step, the resulting array $C \in \mathbb{R}^{T \times l \times r}$ is saved.

To improve upon the first strategy, we use the correlation curves calculated as described above to integrate realistic speckle decorrelation in the simulation by means of a dynamic coherence map, $\mathbf{C}^d \in [0,1]^{T \times H \times W}$, obtained by bi-linear interpolation across the mask of the myocardium.

This coherence map is used to set the relative amplitude between background and myocardial scatterers. To avoid a flickering effect caused by adding and removing background and myocardial scatterers, both sets are superposed in the myocardium. The density of scatterers in the myocardium is therefore double, but their total BSC amplitude will be equivalent. 

At each time step, the myocardial and background scatterers coefficients are set to $\text{BSC}_{m,t}^i = BSC_{m,ES}^i \cdot \mathbf{C}^d_{t, x^i_t,z_t^i}$ and $\text{BSC}_{b,t}^i  = BSC_{b,t}^i \cdot (1 - \mathbf{C}^d_{t, x^i_t,z_t^i})$, respectively. This change in relative amplitude between myocardial and background scatterers causes time-varying speckle correlation in the resulting simulation.

\subsection{Strategy 2: Refinement}

The current improved simulation pipeline still has limitations. Other factors than the coherent and incoherent scatterers can impact speckle correlation in the output simulation. Among other factors, the intensity ratio between the scatterers at the ES instant and the intensity at other time steps can increase speckle correlation in what should be a low correlation region. For this reason, we propose a refinement step. 

To do so, we first do a simulation with strategy 2. We then compute the correlation curves for the simulated sequence at the output of our pipeline: $C_{sim}$. The difference between the desired curves $C$ and the obtained curves can be calculated pointwise $C - C_{sim}$. A new set of input curves can be computed using the following relation: $C^{'} = C + a \cdot (C - C_{sim})$. With this new set of input curves, a new simulation can be generated to have more accurate speckle correlation behavior. We empirically set $a$ to 2.

\section{Results}

\begin{figure*}[tp]
    \centering
    \includegraphics[width=1\linewidth]{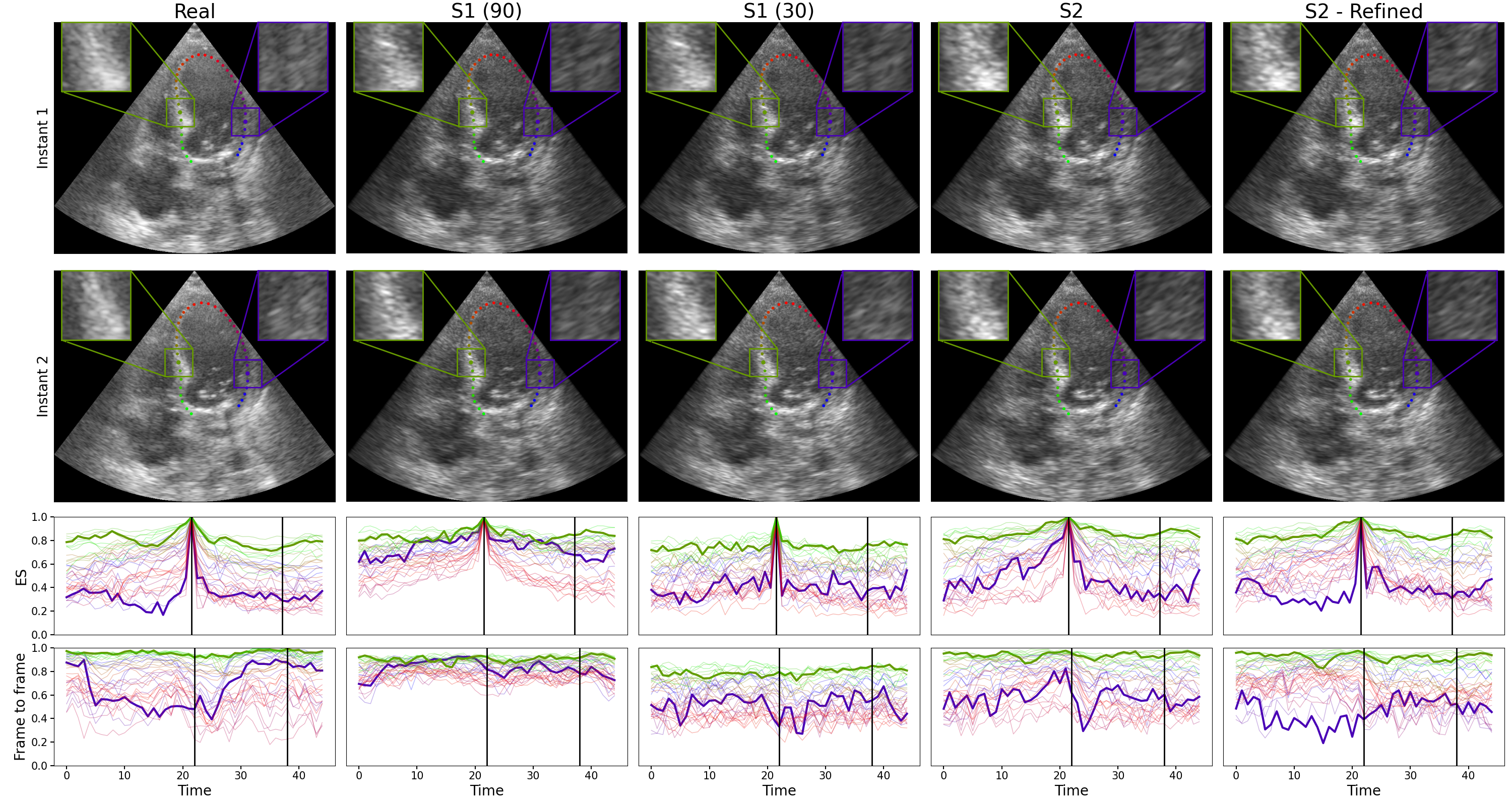}
    \caption{Example of two distinct frames from real and simulated sequences. Correlation curves with ES reference and frame-to-frame reference are shown in the last two rows for the 36 center line points. The highlighted curves represent the points in the zoom regions of the images. Black vertical lines indicate the first and second instants.}
    \label{fig:example}
\end{figure*}




\begin{table}
\centering
\caption{Comparison of correlation curves between simulated and real videos. Strategy 1 is simulated with different probability values $p$.}
\label{tab:table1}
\begin{tabular}{l cc cc}
\toprule
\multirow{2}{*}{\begin{tabular}{c}Simulation \\Type ($p$)\end{tabular}} &
\multicolumn{2}{c}{Frame to Frame correlation} &
\multicolumn{2}{c}{ES correlation} \\
\cmidrule(lr){2-3}\cmidrule(lr){4-5}

& Average corr. & MAE$\downarrow$ & Average corr. & MAE$\downarrow$ \\
\midrule
REAL                                & 0.748 & --    & 0.544 & --    \\
\midrule
S1 (100)                            & 0.849 & 0.117          & 0.700 & 0.163 \\
S1 (90) \cite{evain_camusyn}        & 0.814 & 0.104          & 0.681 & 0.149 \\
S1 (70)                             & 0.768 & 0.093          & 0.646 & 0.120 \\
S1 (50)                             & 0.735 & 0.091          & 0.616 & 0.100 \\
S1 (30)                             & 0.713 & 0.097          & 0.594 & 0.092 \\
S2                                  & 0.762 & 0.085          & 0.643 & 0.119 \\
S2 - Refined                        & 0.731 & \textbf{0.070} & 0.602 & \textbf{0.077} \\


\bottomrule
\end{tabular}

\end{table}

\begin{figure}
    \centering
    \includegraphics[width=1\linewidth]{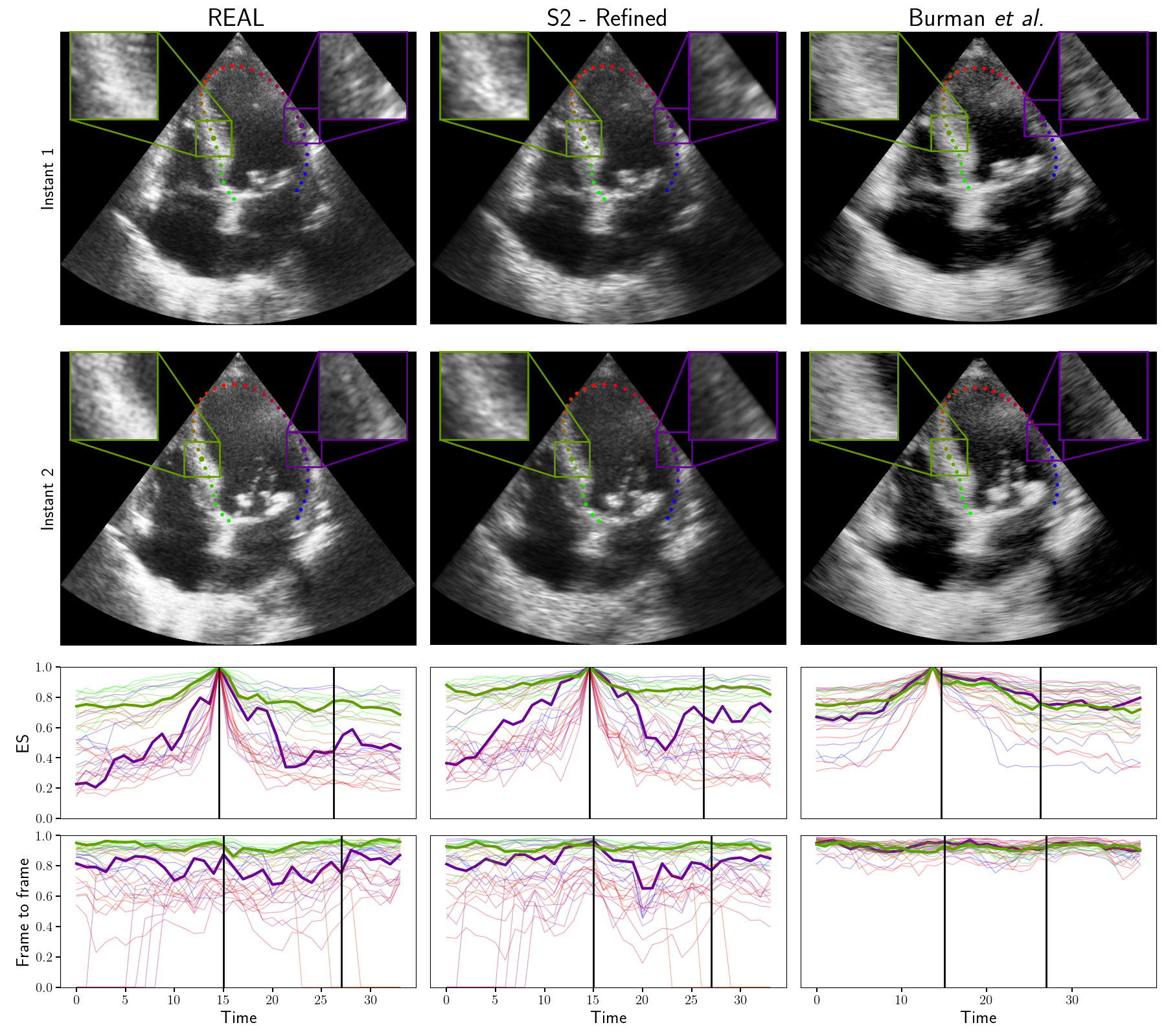}
    \caption{Comparison of refined strategy 2 with simulation from~\cite{burman_dataset}. 
    Display similar to Fig. \ref{fig:example}.}
    \label{fig:example2}
\end{figure}


We conduct an evaluation on a subset of 98 patients from the CAMUS dataset~\cite{leclerc_camus}. For each patient, a full cardiac cycle of the four-chamber view was manually segmented. This segmentation was used to produce the reference motion mesh. We simulated
according to 
the first strategy (\textbf{S1}) with different values of $p$ as well as strategies 2 with and without refinement; \textbf{S2}, \textbf{S2-Refined}. We report the mean absolute error (MAE) between the curves from the real video and simulated video averaged across all 98 sequences. We report the MAE for correlation curves with the reference frame set to ES (as is used for strategy 2). We also compare frame-to-frame correlation 
curves where the reference window is taken in the previous frame, as some optical flow methods work in image pairs rather than full videos. We also compute the average correlation across the sequence as an indicator of the global correlation level and how it compares to real data. 

Results are shown in Table~\ref{tab:table1}. The best results are obtained with strategy \textbf{S2 - Refined}, which outperforms \textbf{S2} and the best version of strategy 1: \textbf{S1-30}. The results show that decreasing the value of $p$ decreases the MAE and average correlation. However, reaching \textbf{S1-30}, the frame-to-frame MAE increases, and the average correlation level is lower than the real data, indicating the correlation is too low. Qualitative results in Fig.~\ref{fig:example} confirm the \textbf{S1-30} simulation has too little correlation. It also shows that both \textbf{S1-90} and \textbf{S1-30} have very constant correlation throughout the video, but at different levels. On the other hand, both versions of strategy 2 have lower correlation but preserve high correlation areas by having more correlation variability both in time and space (across different curves). 

We also compare our proposed simulation with simulations from the dataset proposed by Burman \textit{et al.}~\cite{burman_dataset}. Their dataset is also based on the CAMUS dataset, but on a different subset of patients, precluding a strict quantitative comparison. A qualitative comparison is shown in Fig.~\ref{fig:example2} for one patient present in both datasets, showing that our proposed dataset has much more realistic speckle correlation patterns.

\section{Conclusion}

In this work, we introduced a novel ultrasound simulation framework that explicitly accounts for speckle decorrelation, a key phenomenon often ignored in existing pipelines, but often encountered on real data from clinical routine.
By deriving dynamic coherence maps from real ultrasound data and integrating them into the scatterer model, our approach produces simulated sequences that better capture the spatiotemporal variability of speckle patterns observed in clinical acquisitions. Evaluation on the CAMUS dataset demonstrated that our method achieves lower correlation errors compared to baseline simulations, with improved alignment to real decorrelation behavior across both time and spatial regions of the myocardium. 

Future work could expand on evaluation by introducing better metrics for comparing correlation between videos than MAE on correlation curves and assessing the impact of our simulations on the training and generalization of state-of-the-art motion tracking networks. 


\bibliographystyle{IEEEtran}
\bibliography{main} 


\end{document}